\documentclass[%
 reprint,longbibliography,preprintnumbers,
nofootinbib,
 amsmath,amssymb,
 aps
]{revtex4-1}
\pdfoutput=1
\usepackage{graphicx}
\usepackage[utf8]{inputenc}
\usepackage{flushend}
\usepackage{dcolumn}
\usepackage{bm}
\usepackage{balance}

\usepackage[normalem]{ulem}

\usepackage[colorlinks = true,
            linkcolor = blue,
            urlcolor  = blue,
            citecolor =green,
            anchorcolor = blue]{hyperref}
\usepackage{verbatim}
\usepackage{color,ulem}
\usepackage[english]{babel}

\usepackage[utf8]{inputenc}
\input Starburst.fd
\newcommand*\initfamily{\usefont{U}{Starburst}{xl}{n}}\initfamily

\newcommand{\beq}{\begin{eqnarray}}
\newcommand{\eeq}{\end{eqnarray}}
\usepackage{amsmath}
\usepackage{tikz}
\usetikzlibrary{decorations.pathmorphing}
\usetikzlibrary{shapes.misc}
\tikzset{cross/.style={cross out, draw=black, minimum size=8*(#1-\pgflinewidth), inner sep=0pt, outer sep=0pt},
cross/.default={1pt}}
\usetikzlibrary{patterns,math}
\begin{document}

\title{\Large Reply to Morse and Charbonneau on ``Explicit analytical solution for random close packing in $d=2$ and $d=3$''}

\author{\textbf{Alessio Zaccone}$^{1,2}$}%
 \email{alessio.zaccone@unimi.it}
 
 \vspace{1cm}
 
\affiliation{$^{1}$Department of Physics ``A. Pontremoli'', University of Milan, via Celoria 16,
20133 Milan, Italy.}
\affiliation{$^{2}$Cavendish Laboratory, University of Cambridge, JJ Thomson
Avenue, CB30HE Cambridge, U.K.}


\maketitle
In a quick response to our recent work on an analytical derivation of the random close packing (RCP) density in $d=2$ and $d=3$ based on statistical arguments due to liquid-theory \cite{Zaccone2022}, Morse and Charbonneau (MC) \cite{Charbonneau_Comment} show that our analytical solution works surprisingly well for dimensions $d<6$ ($d=2,3,4,5$), but starts to deviate significantly for higher $d$. Based on this observation they argue that our results are accidental and that the critical weakness of our approach is that it uses liquid state forms ``outside of their domain of validity''.

The work in question \cite{Zaccone2022} established a direct quantitative link between the contact value of $g(r)$, \textit{i.e.} $g(\sigma)$, and the kissing number $z$ using liquid-type theories. The key to achieve this direct relation was to use the splitting between continuous and discrete contributions in the probability density function (pdf) that defines $g(r)$, cfr. Eqs. 5-6 of \cite{Zaccone2022}. Establishing this elusive relationship is arguably the main merit of Ref.~\cite{Zaccone2022}.

MC present an analysis of RCP data for varying numbers of dimensions, $d$, showing that our analytical theory \cite{Zaccone2022} is in good agreement with numerical data at least up to $1/d \approx 0.2$, and deviates significantly beyond this threshold, also showing that for higher $d$ the prediction of the critical density based on the approach in \cite{Zaccone2022} follows an incorrect trend with dimensionality \cite{Charbonneau_Comment}. Based on this observation they arrive at the conclusion that the agreement for $d<6$ is ``accidental'', without providing any evidence in support of this statement other than anecdotal facts based on alternative approaches and analysis that expand around the infinite-dimensional solution for hard spheres \cite{Charbonneau}.

We don't fully agree with their opinion and attitude towards \cite{Zaccone2022}, for the following reasons.

First, we never claimed that an approach based on contact numbers and marginal stability~\cite{Scossa} like ours in \cite{Zaccone2022} would capture the nature of RCP in high dimensional spaces because, for instance, it is known that in $d=8$ there is a large jump in the number of nearest neighbours even at the closest packing \cite{Sloane}. This is because in $d=8$ one gets  $4\, {{8}\choose{2}} + 2^7 = 112+128 = 240$ particles that pack $\sqrt{2}$ away from the origin and from each other (those happen to be the 240 root vectors of the 8-dimensional Euclidean $E_{8}$ Lie group) \cite{Sloane}. We are thus not seeking to compete with the approach of Parisi and co-workers based on replica theory \cite{Charbonneau}, which works well for high-dimensional spaces and becomes exact in $d=\infty$. 



Furthermore, as MC correctly point out, this is not the first attempt at exploiting liquid theories to shed light on RCP, despite being well known that $g(r)$ lacks accuracy upon approaching RCP. Take for instance earlier results published by Kamien and Liu \cite{Kamien} which demonstrated that liquid theories seem to contain the RCP density as a limit, and capture much of the underlying physics, even when one would expect it not to. This, in our opinion, is a fact worth investigating.

It is clear that in our approach \cite{Zaccone2022}, like all simple analytical solutions to complex problems dominated by randomness and disorder, some cancellation of errors must be operative (for example the ``effective boundary condition'' used to determine the constant $g_0$ cannot be perfect, as already pointed out in \cite{Zaccone2022}, and can certainly be improved in future work). A striking, well-known example in soft matter, is Flory's mean-field theory for the equilibrium size of polymer chains in a solvent. This theory is still in use today~\cite{Maritan,Colby,Chaikin} and was successful precisely because of cancellation of errors \cite{deGennes}. Perhaps, following MC's line of thought, we should go as far as saying that Flory theory, too, is fortuitous and its merits should be dismissed?

Finally, MC attribute much importance to results on sphere packings in arbitrary dimensions $d>4$. It should be noted that most of these results may be physically irrelevant. The only space dimensions for which a unified physical theory exists, which is both compatible with Lorentz-invariance and with positive norms upon quantization, and which is able to recover the correct supergravity low-energy limit, is $d=11$ as given by Witten's M-theory \cite{Townsend}.

\bibliographystyle{apsrev4-1}

\bibliography{refs}

\end{document}